\documentclass[onecolumn,floatfix,superscriptaddress,a4paper,
               showpacs,showkeys,nofootinbib,preprint]{revtex4-1}
\usepackage{epsfig}
\usepackage{latexsym}
\usepackage{xspace}
\usepackage{hyperref}
\usepackage[latin2]{inputenc}
\usepackage{indentfirst}
\usepackage{enumerate}

\usepackage{amsmath}
\usepackage{amssymb}
\usepackage[english]{babel}
\usepackage{url}

\newcommand{\eq}[1]{\begin{align} #1 \end{align}}

\begin{document}

\title{Strangeness Production in Light and Intermediate size Nucleus-Nucleus Collisions}

\author{M.I. Gorenstein}
 \affiliation{Bogolyubov Institute for Theoretical Physics, Kiev, Ukraine}
 \affiliation{Frankfurt Institute for Advanced Studies, Frankfurt, Germany}

\author{W. Greiner}
 \affiliation{Frankfurt Institute for Advanced Studies, Frankfurt, Germany}

\author{A. Rustamov}
 \affiliation{Goethe-University Frankfurt, Frankfurt, Germany}

\begin{abstract}
Within the statistical model, the net strangeness
conservation and incomplete total strangeness
equilibration lead to the suppression of strange particle multiplicities.
Furthermore, suppression effects appear to be
stronger in small systems. By treating  the production
of strangeness  within the canonical ensemble formulation
we developed a simple model which allows to predict
the excitation function of $K^+/\pi^+$ ratio in nucleus-nucleus collisions.
In doing so we assumed that different values of $K^+/\pi^+$,
measured in p+p and Pb+Pb interactions at the same collision energy
per nucleon, are driven by the finite size effects only.
These predictions may serve as a
baseline for experimental results from NA61/SHINE
at the CERN SPS and the future CBM experiment at FAIR.

\end{abstract}



\maketitle

\section{Introduction}
The multiplicity of pions per participating nucleon is known to be similar in
nucleus-nucleus (A+A) and in inelastic
proton-proton (p+p) interactions at the same collision energy per
nucleon. This is in line with the Wounded Nucleon Model
\cite{WNM} (WNM) in which the final states in A+A collisions are treated as
a superposition of independent nucleon-nucleon  collisions.
Similar picture emerges from the hadron statistical models
within the grand canonical ensemble (GCE) formulation. At fixed temperature
and chemical potentials all hadron multiplicities are proportional to the system
volume $V$.
Taking $V$ to be proportional to the number of wounded
nucleons $N_W$ in A+A collisions, one restores the WNM results for hadron multiplicities.

Production of strange hadrons appears to be
quite different in p+p and
heavy-ion collisions. In particular, the ratio of $K^+$ to
$\pi^+$ multiplicities is significantly larger in
collisions of heavy ions. It was advocated  to interpret this
{\it strangeness enhancement} as a possible signature for the
quark-gluon plasma creation~\cite{Raf}. A non-monotonic dependence of the
$K^+$ to $\pi^+$ ratio as function of the collision energy
(the {\it horn}) was predicted \cite{GG} as a fingerprint of the
deconfinement phase transition. The predicted behavior was indeed observed by the NA49
Collaboration  in central Pb+Pb collisions \cite{NA49} at the SPS energies
(for more details cf. Ref.~\cite{GGS}). Moreover, these
findings have been recently confirmed by the RHIC and LHC data ~\cite{NA49_Rustamov}.
The experimental data on $K^+/\pi^+$ ratio in p+p and Pb+Pb (Au+Au in the
AGS energy range) collisions are presented in Fig.~\ref{fig1} as
function of the center-of-mass energy of the nucleon pair
$\sqrt{s_{NN}}$ (for details see~\cite{KpiPb} and references therein).

Numbers of strange quarks $N_s$ and antiquarks $N_{\overline{s}}$
in a final state
of p+p or A+A collisions
are equal to each other due to the net strangeness conservation
in strong interactions.
In the SPS energy range strange quarks
are essentially carried by $K^{-}$, $\overline{K^{0}}$ mesons and $\Lambda$ hyperons.
On the other hand, almost all $N_{\overline{s}}$ created in the collision process
are finally revealed in $K^{+}$ and $K^{0}$ particles. For the event averages
one obtains an approximate relation
$<K^{+}>\cong 0.5\, <N_{\bar{s}}>$.
This explains the choice of the $K^+$ multiplicity as an estimator for
the total strangeness~\cite{GGS}.

Conservation of strangeness in large statistical systems can be
treated within the GCE formulation, in which all
hadron multiplicities are proportional to the system volume $V$. In
small systems, however, one has to follow
the canonical ensemble (CE) treatment \cite{ce}.
The multiplicities of (anti)strange hadrons in CE
decrease with decreasing volume
faster than the GCE multiplicities.

A comparison of the statistical model results with hadron multiplicity
data, within both CE and GCE, evidences an incomplete strangeness
equilibration. For reasonable fit of the data one has to introduce the strangeness suppression
factor $\gamma_S$ \cite{gammaS}. Note that in
p+p interactions the $\gamma_S$ factor is smaller than in
central Pb+Pb collisions \cite{becattini}.

In the present study the difference of the $K^+/\pi^+$
ratio in p+p and Pb+Pb collisions is considered
within the CE statistical model as a consequence of
two strangeness suppression effects:
(a) net strangeness conservation and
(b) incomplete total strangeness equilibration.
Our model  assumes that both suppression effects
depend on the system size and collision energy.  Other
physical differences between
statistical systems created in p+p and Pb+Pb collisions
which are not reduced to 'a' and 'b'
are not considered. The finite-size strangeness suppression is
then calculated in terms of two
model parameters which are extracted from existing data on p+p and Pb+Pb
collisions. This opens a possibility to make the model predictions for the
$K^+/\pi^+$ ratio in A+A collisions with light and intermediate
ions. Such estimates are timely in view of experimental program
of the NA61/SHINE at the CERN SPS~\cite{NA61}.
The  NA61/SHINE Collaboration
has already recorded Be+Be data  with projectile momenta of 13{\it A} , 20{\it A},
30{\it A}, 40{\it A}, 80{\it A},
158{\it A}~GeV/c. The energy scans with p+Pb, Ar+Ca and Xe+La collisions
will be completed up to 2016.
In addition, a beam energy scan of Pb+Pb collisions, with much higher
statistics than that performed by the NA49 Collaboration, is
planned. We hope  that the atomic number
dependence of the $K^+/\pi^+$ ratio from p+p to Pb+Pb collisions in the SPS energy range
may reveal new and important physical information.

The Letter is organized as follows. In Section~\ref{Sup} the
strangeness suppression effects in the statistical systems are considered in
the CE formulation.  In
Section~\ref{Pred} the model parameters are extracted from the data on
p+p and Pb+Pb collisions. The model predictions of the $K^+/\pi^+$
ratio for light
and intermediate nucleus-nucleus collisions are calculated.
Finally, Section~\ref{Sum} summarizes the paper.
Appendix A includes details of the calculations.

\section{Strangeness Suppression}\label{Sup}
\begin{figure}[ht!]
\includegraphics[width=0.9\textwidth]{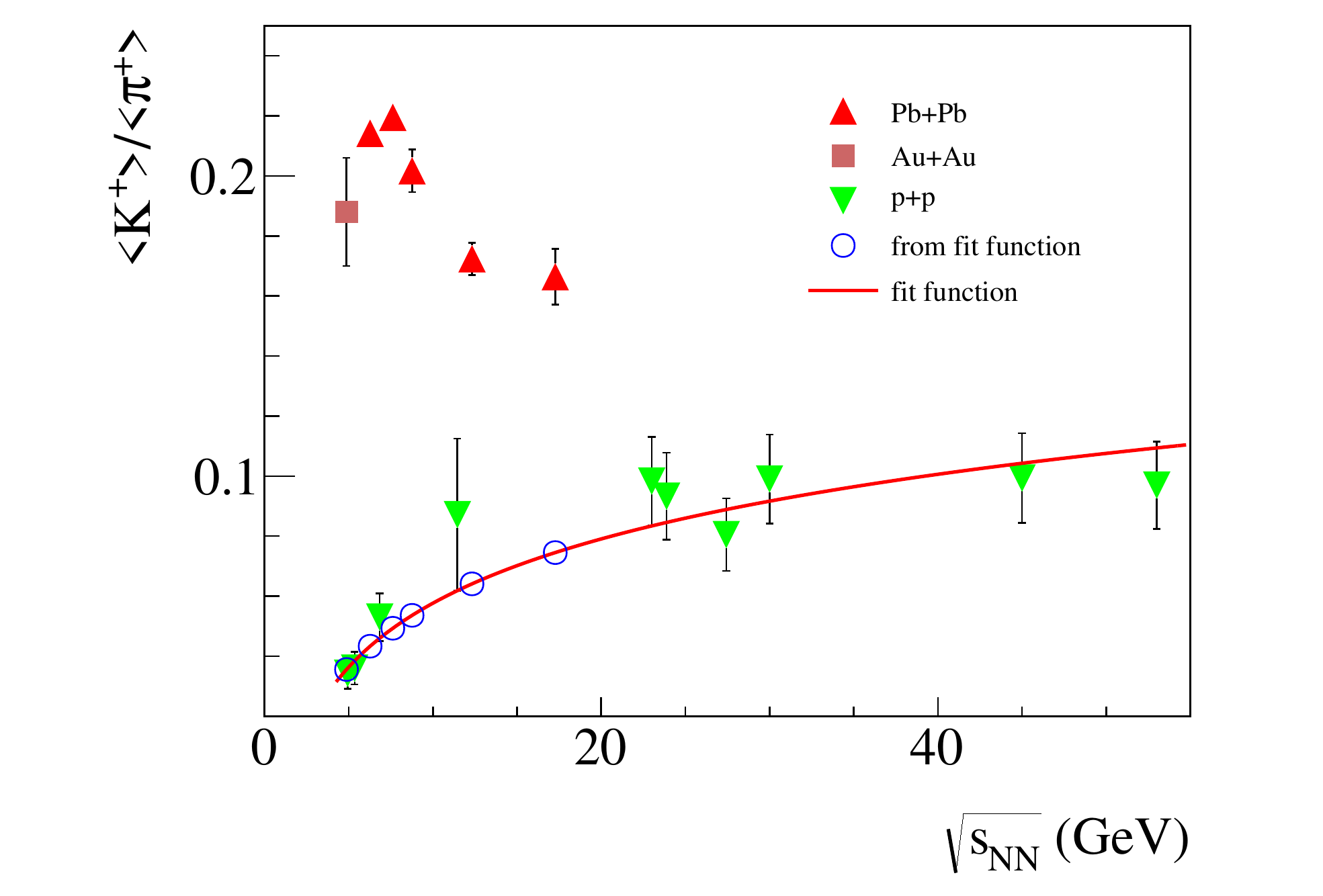}
\caption{(Color Online) The $K^+/\pi^+$ ratio in central Pb+Pb and Au+Au, and
inelastic p+p collisions as a function of the
center-of-mass energy $\sqrt{s_{NN}}$~\cite{KpiPb} . } \label{fig1}
\end{figure}

\begin{figure}[ht!]
\includegraphics[width=0.49\textwidth]{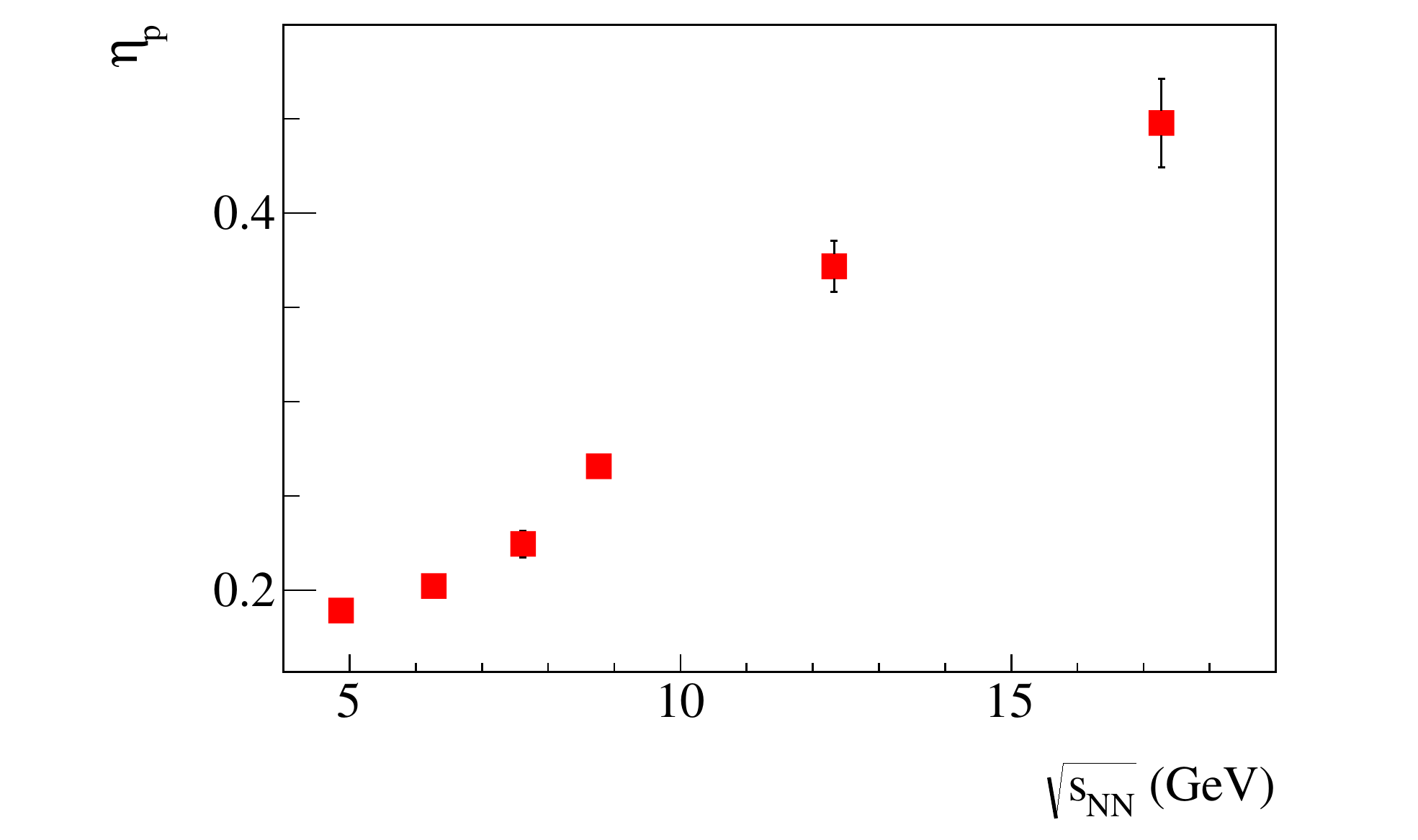}
\includegraphics[width=0.49\textwidth]{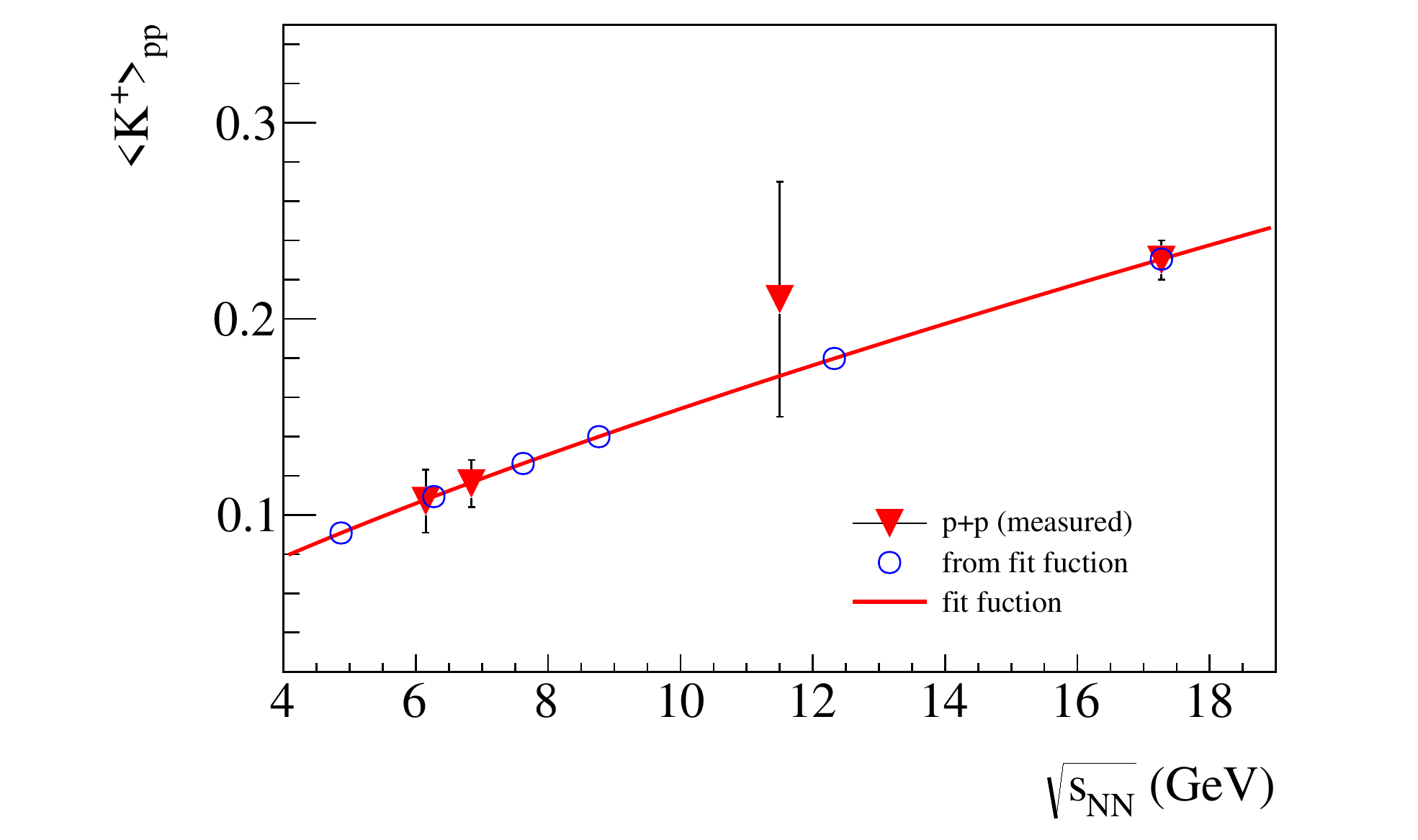}
\caption{ (Color Online) The strangeness suppression factor $\eta_p$ (left panel)
and the multiplicity $\langle K^+\rangle_{pp}$
\cite{Kpp} (right panel) as functions of $\sqrt{s_{NN}}$. }
\label{fig2}
\end{figure}

We first introduce the following notations:
\eq{\label{notations}
 R_p~&\equiv ~\frac{\langle K^+\rangle_{pp} }{\langle \pi^+\rangle_{pp}}~,~~~~~
R_A~ \equiv ~\frac{\langle K^+\rangle_{AA} }{\langle \pi^+\rangle_{AA}}~,~~~~
R_{Pb}~\equiv~\frac{\langle K^+\rangle_{PbPb} }{\langle \pi^+\rangle_{PbPb}}~,~\\
\eta_p~&\equiv ~\frac{R_p}{R_{Pb}}~,~~~~~~~ \eta_A~\equiv ~\frac{R_A}{R_{Pb}}~,
}
where $\langle \ldots \rangle_{pp}$ and $\langle \ldots \rangle_{AA},$
or $\langle \ldots \rangle_{PbPb}$ correspond to the event averages
in inelastic p+p and A+A or Pb+Pb collisions, respectively.
Thereafter the symbol A+A refers to collisions of
light and intermediate size nuclei.
The  data on $R_p$  and  $R_{Pb}$  are presented in
Fig.~\ref{fig1} as function of the center-of-mass energy of a nucleon
pair $\sqrt{s_{NN}}$.
In the left and right panels of Fig.~\ref{fig2}
the energy dependence of $\eta_p$ and  $\langle K^+\rangle_{pp}$ are depicted.
To calculate the $\eta_p$  we use the
$R_{Pb}$ data presented in Fig.~\ref{fig1}
and a function $a+b\cdot(\sqrt{s_{NN}})^{c}$
fitted to the p+p data and shown by the solid line.
The parameters of the function are:
$a= -\,3.397$, $b=3.384$ and $c=0.009$.

The net strangeness conservation requires  equal number
of strange quarks and antiquarks, $N_s-N_{\overline{s}}=0$,
in each event.
The statistical model calculations take into account
global  conservation of the net strangeness.
In the CE
formulation a zero value of the net strangeness is fixed
in each microscopic state of the statistical system.
In  GCE the
chemical potential regulates only the average value of the
net strangeness, i.e. the net strangeness is not necessarily
vanishing in each microscopic state. Both statistical ensembles become
equivalent in the thermodynamical limit when the system volume goes
to infinity. This is discussed in detail in Appendix A.

The $\pi^+$ multiplicity and the quantity
$z$ (see Appendix A, Eq.~(\ref{z})) can be presented as:
\eq{\label{pi+}
\langle \pi^+\rangle_{ii}~=~V_i\,n_{\pi^+}~,~~~~~~ z_i~=~V_i\,n_s~,
}
where $i$=p, A, or Pb. The  $\langle \pi^+\rangle_{ii}$ and $z_i$
correspond, respectively, to the GCE $\pi^+$
multiplicity and $\langle N_s\rangle_{gce}=\langle N_{\overline{s}}\rangle_{gce}$
in $i+i$ collisions.
Note that strange (anti)quark multiplicity $\langle N_s\rangle_{gce}$
corresponds to the complete strangeness equilibration
and does not yet take into account the CE suppression effects.
We assume that the values of the pion number
density $n_{\pi^+}=\langle \pi^+\rangle/V$  and
the strange (anti)quark number density
$n_s=\langle N_s\rangle_{gce}/V$
are not sensitive to the type of reactions,
i.e. they have the same values in p+p, A+A, and Pb+Pb
collisions at the same collision energy.
The volumes $V_i$ are, however, different
in each of these $i+i$ reactions, and
they are assumed to be proportional
to the number of wounded nucleons $N_W$
($N_W=2$ in inelastic p+p collisions).
The GCE formulation will be adopted
for pion multiplicity in all types of $i+i$ collisions.
The total number of negatively charged particles is larger than one
(even in p+p collisions)  at the SPS energies. 
Therefore, the CE effects of electric
charge conservation are small and can be neglected.
To calculate $\langle N_s\rangle=\langle N_{\overline{s}}\rangle $
both the CE effects and the incomplete strangeness
equilibration are considered. This is discussed
in Appendix A (see Eq.~(\ref{Ns-CE})).
For the $K^+$ multiplicity it then follows:
\eq{\label{K-mult}
 \langle K^+\rangle_{ii}~&=~\frac{1}{2}\,
\gamma_S^i\,z_i~\frac{I_1(2\gamma_S^i\,z_i)}{I_0(2\gamma_S^i\,z_i)}~,
}
where the relation $\langle K^+\rangle\cong 0.5 N_{\overline{s}}$ has
been used.

Finally, we obtain the following expressions for $\langle K^+\rangle_{pp}$
and $\eta_p$ in p+p collisions:
\eq{\label{Kmult}
 \langle K^+\rangle_{pp}~&=~\frac{1}{2}\,
\gamma_S^p\,z_p~\frac{I_1(2\gamma_S^p\,z_p)}{I_0(2\gamma_S^p\,z_p)},\\
 \eta_p ~&
 =~
 \frac{\gamma_S^p}{\gamma_S^{Pb}}~
\frac{I_1(2\gamma_S^p\,z_p)}{I_0(2\gamma_S^p\,z_p)}~.\label{eta}
}
The above equations assume:
(i) the same $n_s$ and $n_{\pi^+}$ GCE values of the particle number densities, as defined in
Eq.~(\ref{pi+})
in p+p,  A+A, and Pb+Pb collisions;  (ii) the incomplete strangeness
equilibration regulated by $\gamma_S^i$ in $i+i$ collisions ($i=p$, A, and Pb);
(iii) the relation $I_1/I_0 \cong 1$ is adopted in central Pb+Pb collisions,
as  $\gamma_S^{Pb}\,z_{Pb}\gg 1$.

\section{Predictions for Light Ion Collisions}\label{Pred}
\begin{figure}[ht!]
\includegraphics[width=0.49\textwidth]{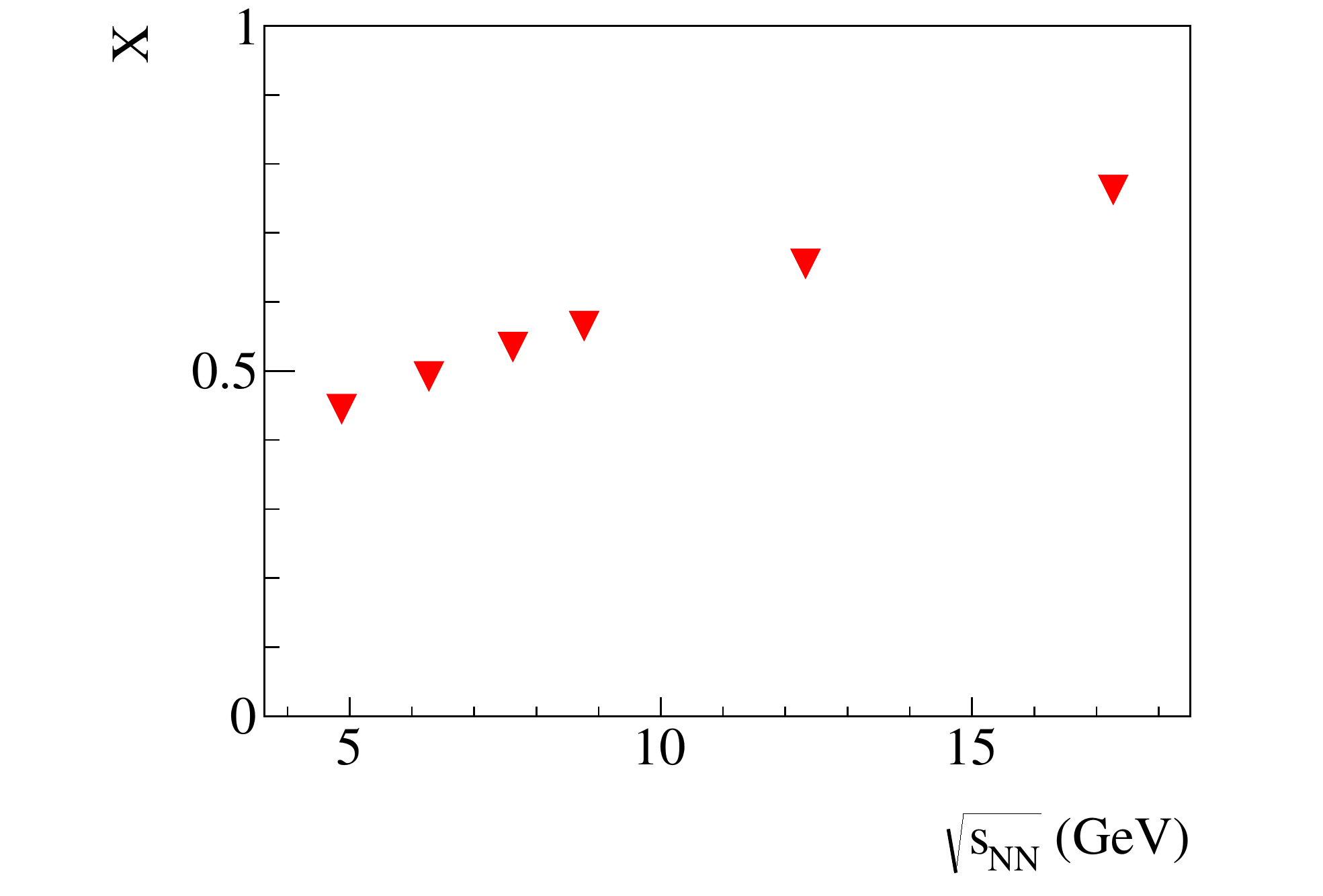}
\includegraphics[width=0.49\textwidth]{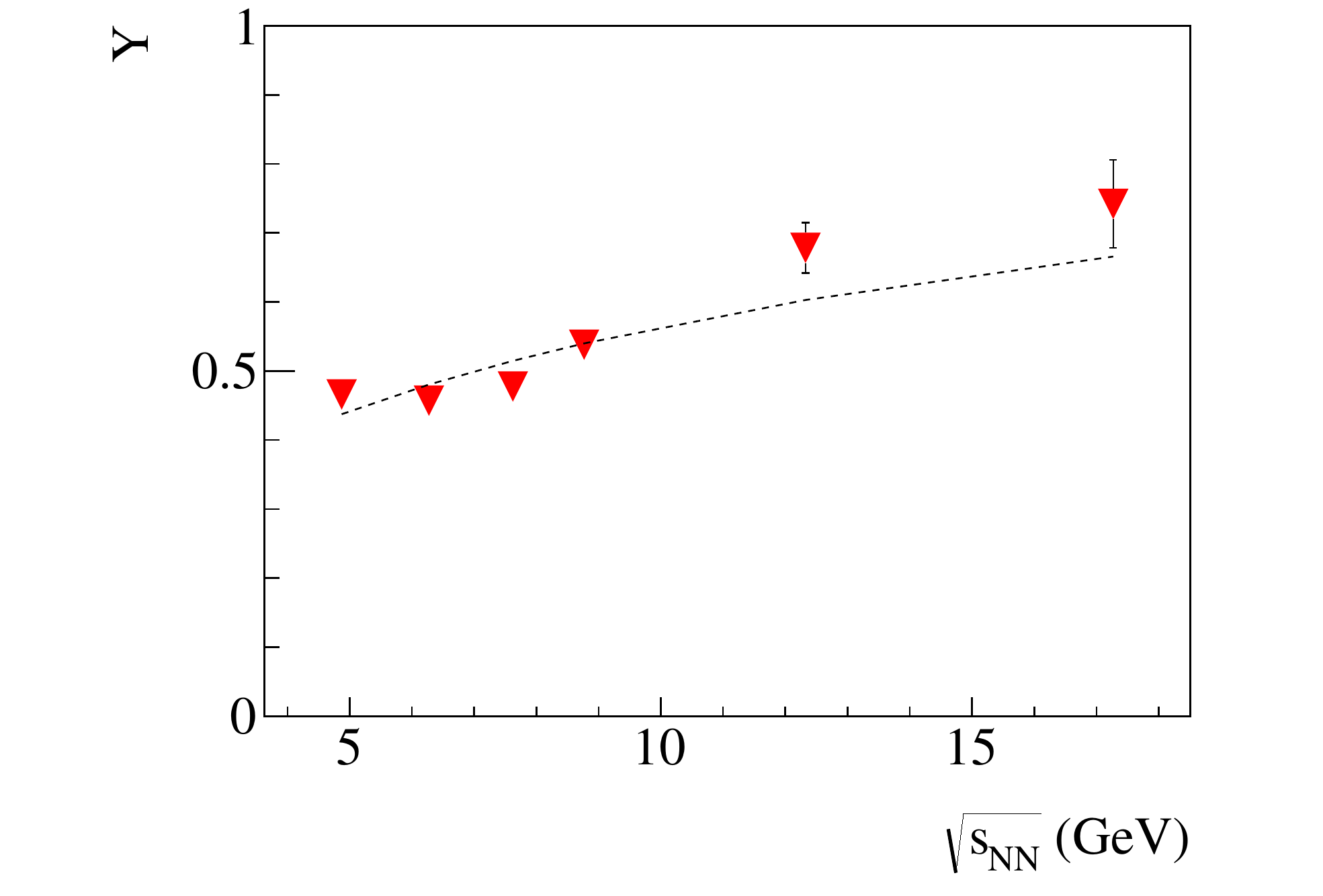}
\caption{(Color Online) The energy dependence
of the solutions $X$ and $Y$ of Eqs.~(\ref{Kmult1}),~(\ref{eta1}).
Left panel: $X=\gamma^p_S z_p$. Right panel:
$Y=\gamma_S^p/\gamma_S^{Pb}$. The dashed line represents the fit with
the Eq.~(\ref{gammaSA}), yielding $\alpha=1.015$ and $\beta$=0.189.}
\label{fig3}
\end{figure}

The left-hand-sides of Eqs.~(\ref{Kmult}) and (\ref{eta}) involve quantities which
have  been experimentally measured. The energy dependences of $\langle K^+\rangle_{pp}$ and $\eta_p$
are shown in
Fig.~\ref{fig2}. For the $\langle K^+\rangle_{pp}$ we used
the fit function $a\cdot(\sqrt{s_{NN}})^{b}$ with $a$=0.028
and $b$=0.736 presented by the solid line in the right panel of Fig.~\ref{fig2}.
All in all there are 3 unknowns, $\gamma_S^p$, $z_p$, and $\gamma_S^{Pb}$, entering to
the right-hand-sides of  Eqs.~(\ref{Kmult}) and (\ref{eta}). However,
they can be combined as
\eq{\label{XY}
X~=~\gamma_S^p\,z_p~,~~~~~ Y~=~\gamma_S^p/\gamma_S^{Pb}~.
}
Together with Eq.~(\ref{XY}),  Eqs.~(\ref{Kmult}) and (\ref{eta}) represent
the system of two equations with two
unknown quantities:
\eq{\label{Kmult1}
 \langle K^+\rangle_{pp}~&=~\frac{1}{2}\,
X~\frac{I_1(2X)}{I_0(2X},\\
 \eta_p ~&
 =~Y~\frac{I_1(2X)}{I_0(2X)}~.\label{eta1}
}
The solution of the
transcendental Eq.~(\ref{Kmult1}),
$X=X(\sqrt{s_{NN}})$, is shown in the left panel of Fig.~\ref{fig3}.
On the other hand, Eq.~(\ref{eta1}) gives
the value of $Y=\eta_p I_0(2X)/I_1(2X)$~
presented in the right panel of Fig.~\ref{fig3}.

\begin{figure}[ht!]
\includegraphics[width=0.8\textwidth]{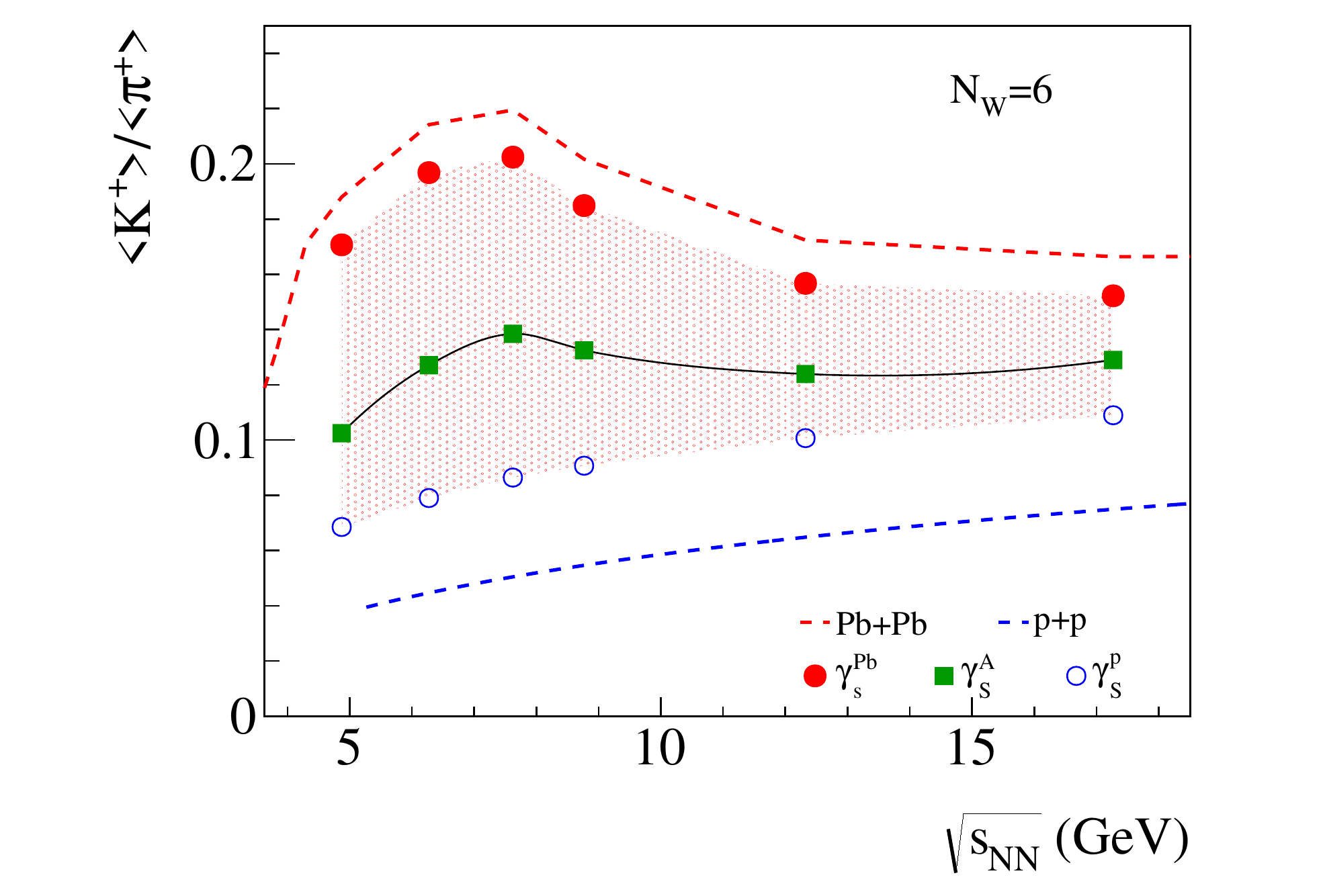}
\includegraphics[width=0.8\textwidth]{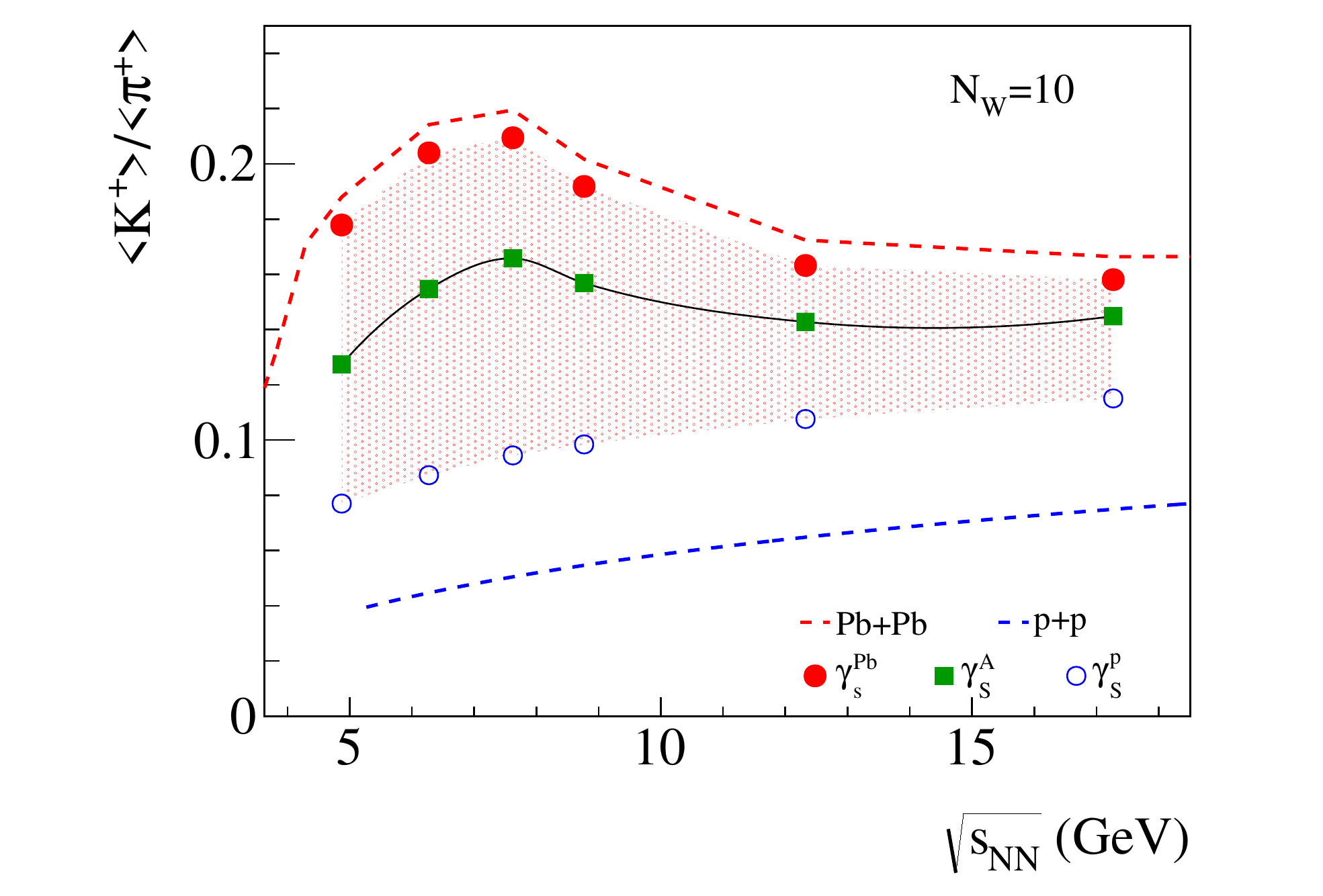}
\caption{(Color online) The energy dependence of  $R_A$ as calculated by Eq.(\ref{KpiA}) for $N_W = 6$ (upper panel) and  $N_W = 10 $ (lower panel) are presented
with green boxes. The dashed lines represent measurements in p+p 
(lower line) and Pb+Pb (upper line) collisions. 
The lower (open circles)  and upper (full circles) limits
are calculated using  Eqs.~(\ref{low}) and~(\ref{up}), respectively.}
\label{fig4}
\end{figure}

\begin{figure}[ht!]
\includegraphics[width=0.8\textwidth]{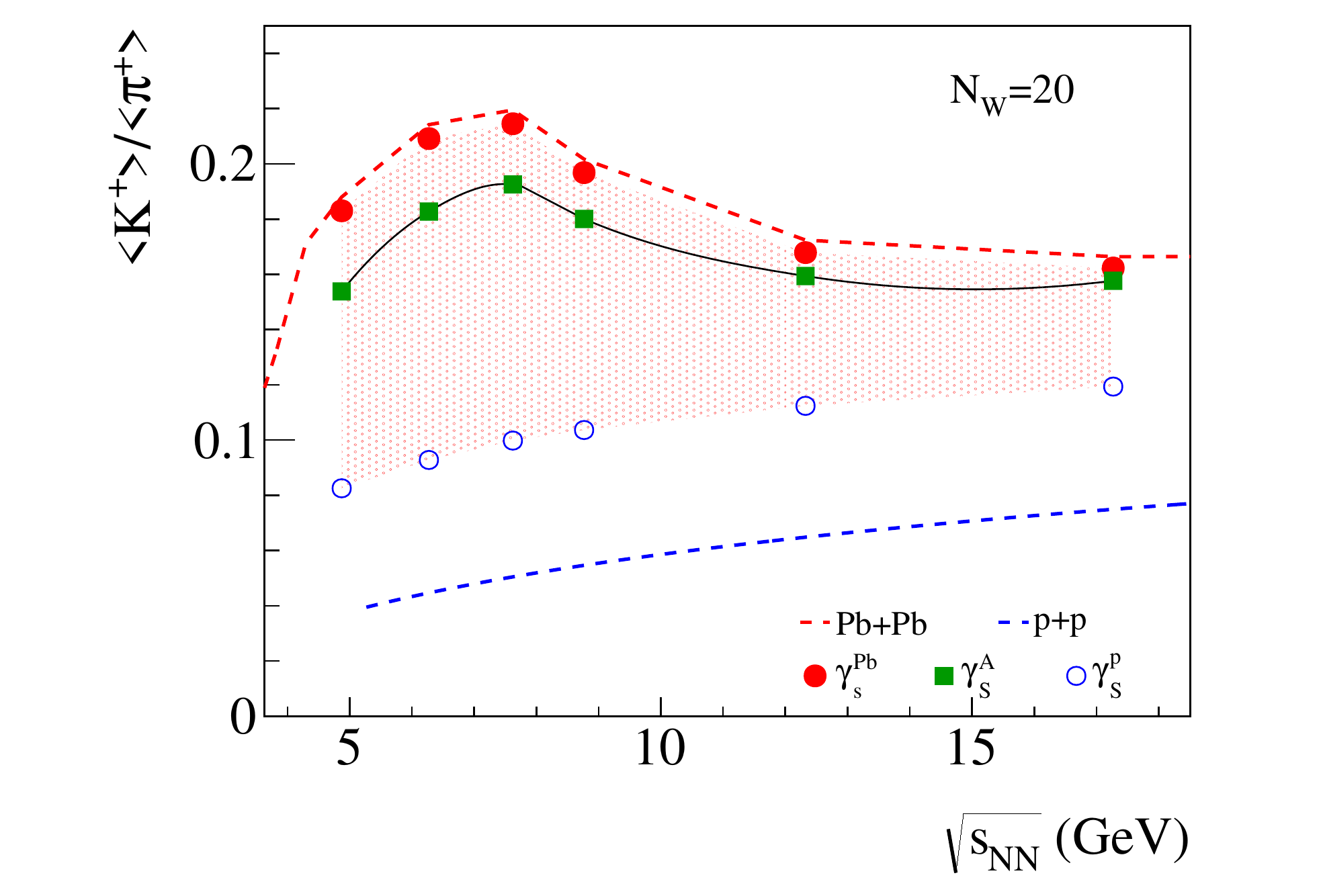}
\includegraphics[width=0.8\textwidth]{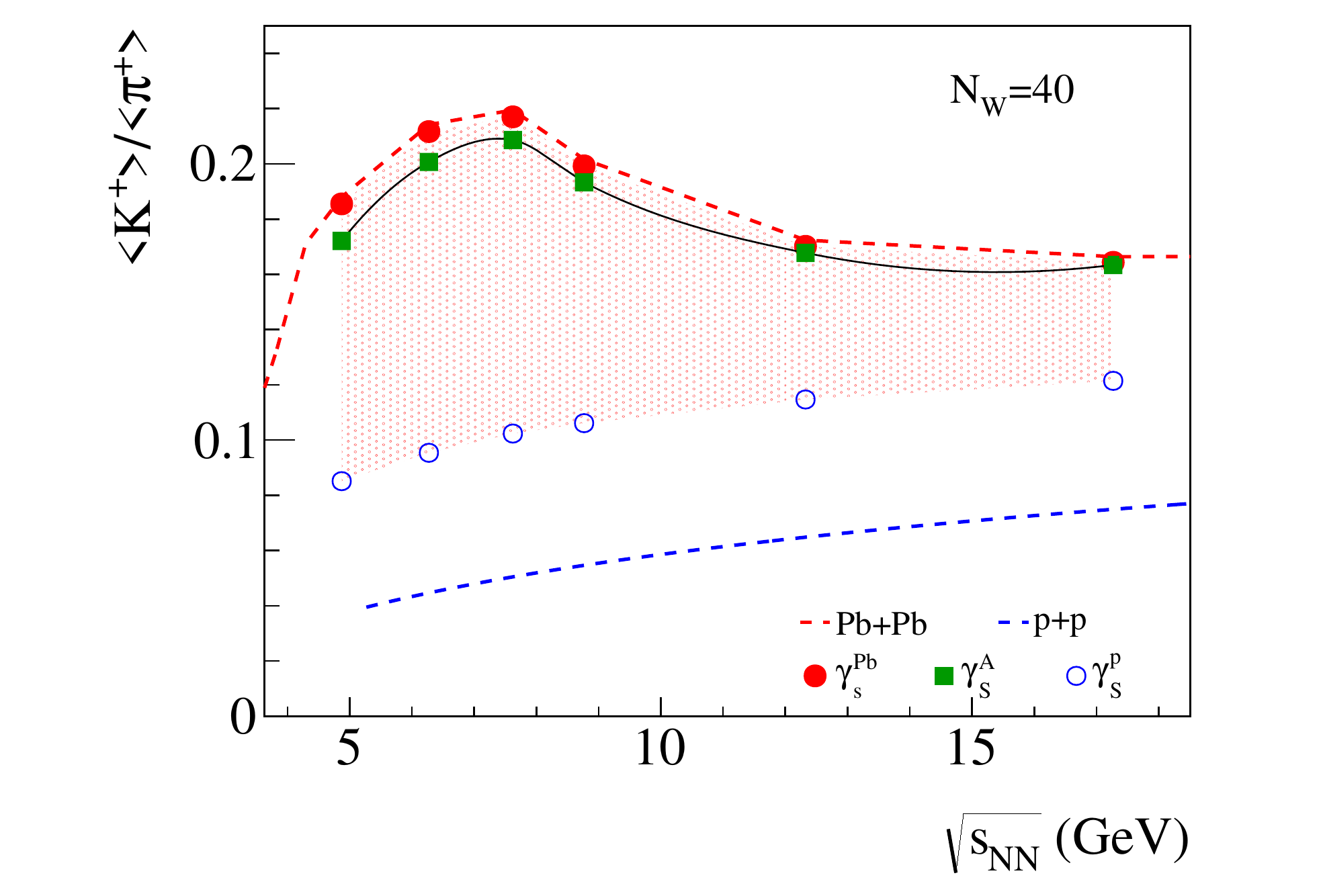}
\caption{ (Color Online) 
The same as in Fig.~\ref{fig4} but for $N_W = 20$ (upper panel) and  $N_W = 40 $ (lower panel).
}
\label{fig5}
\end{figure}

Assuming now $z_A=z_p\cdot N_W/2$, where $N_W$ is the average number
of wounded nucleons in A+A collisions,  one can calculate
the $K^+$ to $\pi^+$ ratio as:
\eq{\label{KpiA}
R_A\equiv
\frac{ \langle K^+\rangle_{AA}}
{\langle \pi^+\rangle_{AA}}=
R_{Pb} \times \frac{\gamma_S^A}{\gamma_S^{Pb}}
\cdot\frac{I_1(2\gamma_S^A z_p \cdot N_W/2)}
{I_0(2\gamma_S^A z_p\cdot N_W/2)}~
=
R_{Pb} \times \frac{\gamma_S^A}{\gamma_S^{Pb}}
\cdot\frac{I_1[(\gamma_S^A/\gamma_S^p) \,X \cdot N_W]}
{I_0[(\gamma_S^A/\gamma_S^p)\, X \cdot N_W]}~.
}
%

\begin{figure}[ht!]
\includegraphics[width=0.9\textwidth]{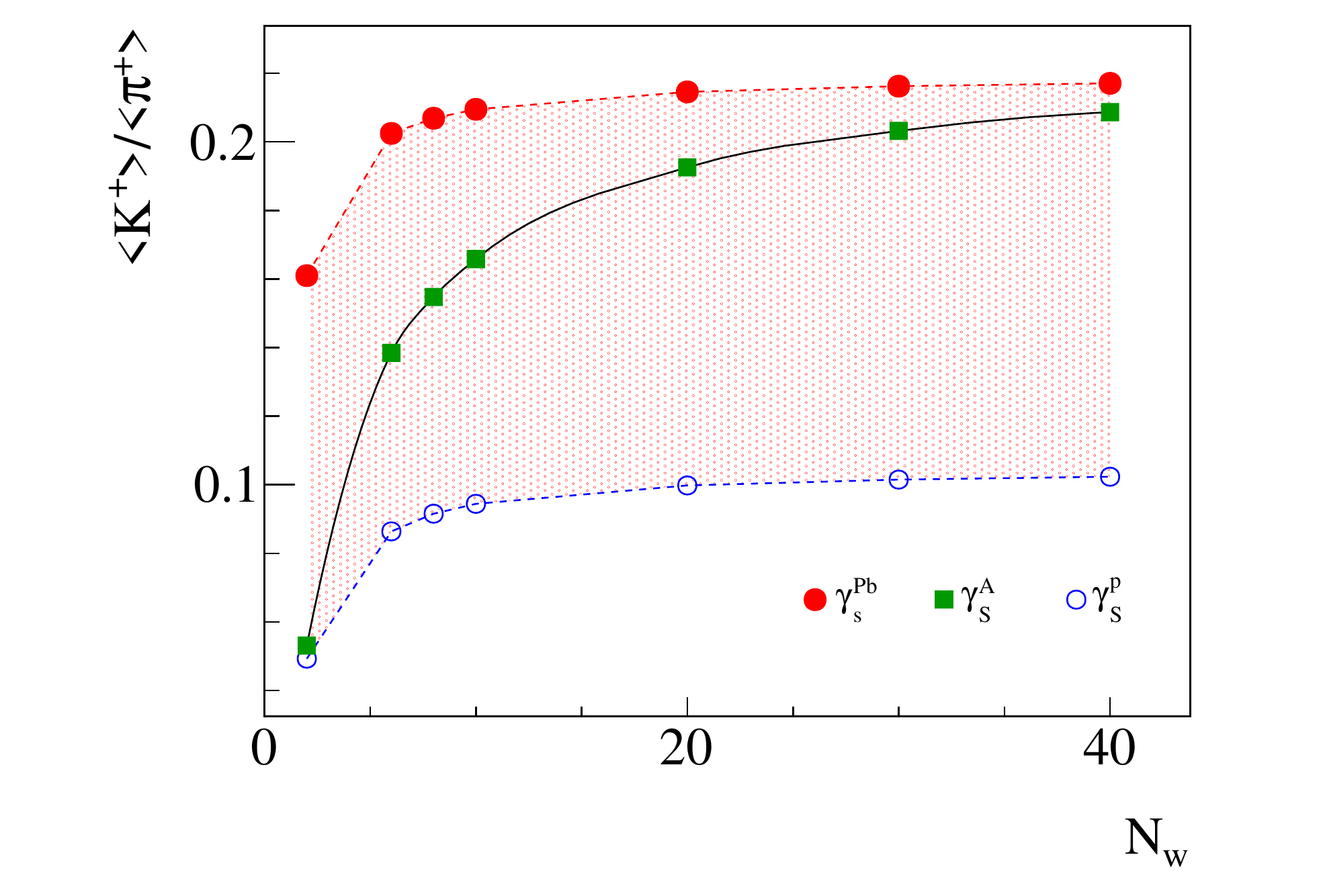}
\caption{(Color Online) The dependence of $\langle K^+\rangle/\langle \pi^+\rangle$ on the number of wounded
nucleons $N_W$ in A+A collisions at fixed energy of $\sqrt{s_{NN}}=7.6$~GeV are presented
with green boxes. The lower (open circles)  and upper (full circles) limits
are calculated using  Eqs.~(\ref{low}) and~(\ref{up}), respectively. }
\label{fig6}
\end{figure}

Next, following the prescription of Ref.~\cite{BG}, we used the following expression for the dependence of $\gamma_S^A$ on $N_W$ and $\sqrt{s_{NN}}$:

\eq{\label{gammaSA}
\gamma_S^A~=~1~-~\alpha\,\exp\Big[-\,\beta\,\sqrt{N_W\,\sqrt{s_{NN}}}\,\Big]
}
with $\alpha$=1.015 and $\beta$ =0.189, which were obtained by fitting the 
$\gamma_S^p$/$\gamma_S^{Pb}$  ratio (see the right panel of Fig.~\ref{fig3}).

Furthermore, taking $\gamma_S^A=\gamma_S^p$ and $\gamma_S^A=\gamma_S^{Pb}$,
we obtain the lower ($R_A^{\rm low}$) and upper ($R_A^{\rm up}$) limits for $R_A$ defined in Eq.~(\ref{KpiA}):
%
\eq{
R_A^{\rm low}~&=~
R_{Pb} \times Y \cdot
\frac{I_1[X \cdot N_W]}
{I_0[X \cdot N_W]}~,\label{low}\\
R_A^{\rm up}~&=~
R_{Pb} \times
\frac{I_1[Y^{-1} \,X \cdot N_W]}
{I_0[Y^{-1}\, X \cdot N_W]}~.\label{up}
}

In Fig.~\ref{fig4} and Fig.~\ref{fig5} the energy dependence of $R_{A}$ for A+A collisions with different numbers of wounded nucleons $N_W$ are presented.
The green boxes are calculated using  Eqs.~(\ref{KpiA}) and (\ref{gammaSA}).
The lower and upper dashed lines correspond to the $K^+/\pi^+$ ratios
in p+p and Pb+Pb collisions, respectively. The open and full circles
are calculated using Eqs.~(\ref{low}) and~(\ref{up}), correspondingly.

%
%

%

In Fig.~\ref{fig6} we illustrate with green boxes the system size dependence (expressed in terms of wounded nucleons)
of the $\langle K^+\rangle/\langle \pi^+\rangle$ ratio at fixed energy of $\sqrt{s_{NN}}=7.6$~GeV.
The upper limit (full circles) corresponds to $\gamma_S^A=\gamma_S^{Pb}$ and the lower  limit (open circles)
to $\gamma_S^A=\gamma_S^p$. Interestingly, the $\langle K^+\rangle/\langle \pi^+\rangle$
ratio becomes approximately independent of the number of wounded nucleons
for $N_W > 40$.

%
%
%
%


\section{Summary}\label{Sum}
In summary, the $K^+/\pi^+$
ratio in p+p and Pb+Pb collisions is considered
within the statistical model. The model takes into account
the net strangeness conservation within the canonical ensemble
formulation and the incomplete total strangeness equilibration
regulated by the parameter $\gamma_S$.
Both effects
are assumed to depend on the system size only.
The two model parameters are
extracted from the existing data in p+p and Pb+Pb
collisions.
We present the model estimates
for the lower and upper limits of $R_A$, defined in Eq.~(\ref{KpiA}), for A+A collisions which
correspond to $\gamma_S^A=\gamma_S^p$ and $\gamma_S^A=\gamma_S^{Pb}$,
respectively.
%
Assuming a functional dependence of $\gamma_S^A$ on $N_W$ and $\sqrt{s_{NN}}$
in the form of Eq.~(\ref{gammaSA}) we managed to make definite predictions for the $K^+/\pi^+$ ratio in collisions of light and intermediate
nuclei at the SPS energy region.
We hope that our estimates will be helpful for
the NA61 SHINE program
with collisions between light and intermediate size nuclei.
In particular,
the deviations
of the future experimental results from our predictions, if there will be any, will clearly underline important physics differences between p+p and A+A collisions.

\vspace{0.3 cm}
\begin{acknowledgments}
We would like to thank Marek Ga\'zdzicki,
Francesco Becattini and Herbert Str\"{o}bele for
fruitful discussions and comments.
The work of M.I.G. was supported by the Program of
Fundamental Research of the Department of Physics and
Astronomy of NAS, Ukraine, and by the State Agency of Fundamental Research of Ukraine,
Grant F58/04. A.R. gratefully acknowledges
the support by the German Research Foundation (DFG Grant No. GA 1480/2.1).

\end{acknowledgments}

\section*{Appendix}
The GCE partition function for strange quarks and antiquarks reads
\eq{\label{GCE-PF}
Z_{gce}(T,V,\gamma_S;\lambda,\overline{\lambda})~&=~\sum_{N_s=0}^{\infty}\sum_{N_{\overline{s}}=0}^\infty
\frac{(\gamma_S\,\lambda\,z)^{N_{s}}}{N_s!}\,
\frac{(\gamma_S\,\overline{\lambda}\,z)^{N_{\overline{s}}}}{N_{\overline{s}}!}~\nonumber \\
&=
~\exp\Big(\gamma_S\,\lambda\,z~+~\gamma_S\,\overline{\lambda}\,z\Big)~
\rightarrow~\exp\Big( 2\gamma_S\,z\Big)~,
}
where the quantity $z$ is the so-called one-particle partition function
\eq{
z~=~\frac{V}{\pi^2}~Tm_s^2\,
K_2\Big(\frac{m_s}{T}\Big)~\equiv~V\cdot n_s~. \label{z}
}
In Eqs.~(\ref{GCE-PF}),~(\ref{z}), $V$ and $T$ are the system volume and temperature,
respectively,  $m_s$ is the mass of strange (anti)quark and
$K_2$ is the modified Bessel
function.
Furthermore, the Boltzmann
approximation is used because the quantum statistics effects
are negligible.
The $\lambda$ and $\overline{\lambda}$ in Eq.~(\ref{GCE-PF})
are auxiliary parameters introduced
to calculate $N_s$ and $N_{\overline{s}}$ averages:
\eq{\label{NsGCE}
\langle N_s\rangle_{gce}~=~\Big[\frac{\partial \ln Z_{gce}}
{\partial \lambda}\Big]_{\lambda=\overline{\lambda}=1}~
=~
\langle N_{\overline{s}}\rangle_{gce}~
=~\Big[\frac{\partial
\ln Z_{gce}}{\partial \overline{\lambda}}\Big]_{\lambda=\overline{\lambda}=1}~
=~\gamma_S\,z~,
}
The parameter $\gamma_S$
regulates the strangeness equilibration \cite{gammaS}. It is used
to fit the average value of the total
strangeness measured by experiments: $\gamma_S<1$
corresponds to an incomplete strangeness equilibration, whereas
$\gamma_S=1$ means a complete chemical equilibrium.

The GCE partition function (see Eq.~(\ref{GCE-PF})) leads to the equal average values of
$N_s$ and $N_{\overline{s}}$.
However, the terms with $N_s\neq N_{\overline{s}}$ contribute to $Z_{gce}$.
On the other hand, the CE partition function requires $N_s=N_{\overline{s}}$ in each microscopic state
of the system:
\eq{\label{CE-PF}
Z_{ce}(T,V,\gamma_S;\lambda,\overline{\lambda})~&=~
\sum_{N_s=0}^{\infty}\sum_{N_{\overline{s}}=0}^\infty
\frac{(\gamma_S\,\lambda\,z)^{N_{s}}}{N_s!}\,
\frac{(\gamma_S\,\overline{\lambda}\,z)^{N_{\overline{s}}}}{N_{\overline{s}}!}~
\delta(N_s-N_{\overline{s}})\nonumber \\
&=~\frac{1}{2\pi}\int_0^{2\pi}d\phi\,
\exp\Big[\gamma_S\,z\left(\lambda e^{i\phi}+\overline{\lambda}
e^{-i\phi}\right)\Big]~\rightarrow~I_0(2\gamma_S\,z)~.
}
The average numbers of strange quarks and
antiquarks become:
\eq{\label{Ns-CE}
\langle  N_s\rangle_{ce} ~=~
\Big[\frac{\partial \ln Z_{ce}}
{\partial \lambda}\Big]_{\lambda=\overline{\lambda}=1}~
=~
\langle  N_{\overline{s}}\rangle_{ce} ~=~
\Big[\frac{\partial \ln Z_{ce}}{\partial
\overline{\lambda}}\Big]_{\lambda=\overline{\lambda}=1}~
=~
\gamma_S\,z~\cdot \frac{I_1(2\gamma_S\,z)}{I_0(2\gamma_S\,z)}~.
}
The ratio of Bessel
functions $I_1$ and $I_0$ in Eq.~(\ref{Ns-CE}) describes the
suppression effect due to  conservation of the net
strangeness in each microscopic state of the CE.
The CE suppression factor $I_1/I_0$ is a function of $\gamma_S\,z$.
Thus, only this quantity defines  the CE effects, the
specific values of $m_s$, $T$, and $V$ are irrelevant.
For $\gamma_S\,z\gg 1$ it follows that $I_1(2\gamma_S\,z)/I_0(2\gamma_S\,z)\cong 1$.
Therefore, for large systems, the CE suppression effects are
negligible, i.e., the CE and GCE multiplicities become identical.


\end{document}